\documentclass[12pt]{article}

\usepackage{xcolor}
\usepackage{url}
\usepackage{latexsym}
\usepackage{amsmath}
\usepackage{booktabs}
\usepackage{amssymb}
\usepackage{relsize}
\usepackage{graphicx}
\usepackage{epsfig}
\usepackage{ccaption}

\usepackage{a4wide}

\usepackage[percent]{overpic}

\newcommand{\be}{\begin{equation}}
\newcommand{\ee}{\end{equation}}
\newcommand{\bea}{\begin{eqnarray}}
\newcommand{\eea}{\end{eqnarray}}
\newcommand{\nn}{\nonumber\\}

\newtheorem{Con}{Conjecture}[section]

\def\ds{\displaystyle}
\def\pd{\partial}
\def\Area{\mathrm{Area}}
\def\za{z_\ast}

\def\m{\mu}
\def\n{\nu}

\def\Om{\Omega}
\def\th{\theta}
\def\p{\phi}
\def\D{\Delta}

\def\L{\mathcal{L}}

\def\e{\varepsilon}
\def\G{\Gamma}
\def\g{\gamma}
\def\la{\lambda}

\def\A{\mathcal{A}}

\def\H{\mathcal{H}}
\def\O{\mathcal{O}}
\def\N{\mathcal{N}}

\def\R{\mathbb{R}}
\def\Z{\mathbb{Z}}

\def\thus{\Rightarrow}
\def\tend{\rightarrow}

\title{\bf Extracting Spacetimes using the AdS/CFT Conjecture: Part II}
\author{Samuel Bilson\footnote{s.c.bilson@dur.ac.uk}\\ \\
{\it Department of Mathematical Sciences,}
\\
{\it Science Laboratories, South Road, Durham DH1 3LE, United Kingdom}}

\begin{document}
\begin{titlepage}
\maketitle

\maketitle

\begin{abstract}
Motivated by the holographic principle, within the context of the AdS/CFT Correspondence in the large t'Hooft limit, we investigate how the geometry of certain highly symmetric bulk spacetimes can be recovered given information of physical quantities in the dual boundary CFT. In particular, we use holographic entanglement entropy proposal (relating the entanglement entropy of certain subsystems on the boundary to the area of static minimal surfaces) to recover the bulk metric using higher dimensional minimal surface probes within a class of static, planar symmetric, asymptotically AdS spacetimes. We find analytic and perturbative expressions for the metric function in terms of the entanglement entropy of straight belt and circular disk subsystems of the boundary theory respectively. Finally, we discuss how such extractions can be generalised.    
\end{abstract}
\thispagestyle{empty}
\end{titlepage}

\section{Introduction}
\label{Intro}
\numberwithin{equation}{section}

Maldecena's AdS/CFT correspondence\cite{Maldacena,MAGOO,Witten1} states that
\begin{Con}
\textit{Type IIB superstring theory with} $AdS_5\times S^5$ \textit{boundary conditions is equivalent to} $\mathcal{N}=4$, \textit{SU(N) Super-Yang-Mills theory (SYM) in} $3+1$ \textit{dimensions}. 
\end{Con}
$\mathcal{N}=4$, SYM in four dimensions is a non-Abelian supersymmetric gauge theory with conformal symmetry. Anti-de Sitter, or $AdS$ space, is a maximally symmetric solution of Einstein's equations with negative cosmological constant. This implies that the AdS/CFT correspondence is a duality between a four dimensional gauge theory and a five dimensional gravitational theory. In some sense, it is useful to think of the gauge theory as ``living on the boundary'' of the bulk AdS spacetime. Infrared effects in AdS space correspond to ultraviolet effects in the boundary theory, known as the UV/IR connection. The fact that the physics in the bulk $AdS$ space can be described by a field theory of one less dimension is one example of the ``holographic principle''. In a quantum gravity theory, all physics within some volume can be described in terms of some theory on the boundary which has one bit of information per Planck area. This holographic bound is the physical interpretation of the UV/IR connection\cite{Hol5}.\\

The idea of a holographic interpretation of a quantum gravity theory in terms of a boundary gauge theory lead Polchinski and Horowitz\cite{adscft2} and others to propose a ``gauge/gravity duality''. This asserts that
\begin{Con}
\textit{Hidden within every non-Abelian gauge theory, even within the weak and strong nuclear interactions, is a theory of quantum gravity}.
\end{Con}
Such a duality can be motivated without the notion of string theory, although one finds string theory is hidden within this description.\\

The fact that gauge theories and string theories are related is not unusual. It had already been noted by t'Hooft\cite{largeN} in 1973 that the large $N$ limit (where $N$ is the number of colours) of certain gauge theories at strong coupling correspond to worldsheet string perturbation theory at weak coupling. It is this strong/weak coupling duality in AdS/CFT that allows us to study the non-perturbative regime of string theory by looking at weakly coupled gauge theories.\\

An immediate application of AdS/CFT was proposed by Witten\cite{Witten1}, where he found a one-to-one correspondence between operators in the field theory and fields propagating in AdS space by equating the generating functional of correlation functions in the CFT to the full partition function of string theory. This allows computation of correlation functions in the gauge theory in terms of supergravity Feynman diagrams. This is an explicit manifestation of the AdS/CFT correspondence and provides us with a dictionary from which one can relate observables in the bulk to observables on the boundary. In particular, since AdS/CFT relates weak and strong coupling, bulk physics with high curvature is related to weakly coupled field theories where one can perform perturbation theory. This has been applied in a process known as holographic renormalization where one can obtain renormalized QFT correlation functions by performing computations on the gravity side of the correspondence\cite{HR}. This has been used to compute such quantities as the expectation value of the boundary CFT stress-energy-momentum tensor associated with a gravitating system in asymptotically anti-de Sitter space\cite{Tmn}.\\

By relating thermodynamic properties on both sides of the correspondence, it was shown\cite{Witten} that $\N=4$ SYM theory in a thermal state is dual to a large mass, AdS-Schwarzschild black hole. This was developed by \cite{Maldacena1} to argue that an eternal black hole in AdS spacetime can be holographically described by two identical copies of the dual CFT associated with the spacetime and an initial entangled state. This allowed a whole swathe of research on trying to understand the basic non-perturbative aspects on both sides of the correspondence. Characteristic properties of confinement and asymptotic freedom in QCD have been investigated using the AdS/QCD correspondence\cite{AdS/QCD1,AdS/QCD2,AdS/QCD3,AdS/QCD4}, where one can perform calculations in a weakly coupled string theory. More recently, superfluidity and superconductivity in, for example, strongly coupled plasmas have been studied by applying AdS/CFT to condensed matter systems, known as AdS/CMT\cite{AdS/CMT1,AdS/CMT2,AdS/CMT3,AdS/CMT4}. On the gravity side, properties such as causal structure, event horizons and singularities could equally well be understood using strongly coupled gauge theories. In particular, one might hope to better understand the physics of black holes, including the description of physics behind the horizon.\\

By identifying field theoretic observables, in particular boundary correlation functions, work was done on identifying signals of the bulk curvature singularity\cite{Fid,Bal,Louko,Kraus,Fest,Freiv}. It was observed that signals of black hole singularities in the bulk can be identified by considering nearly null spacelike probes. These provide a large contribution to the boundary field theory correlation function, giving rise to ``light-cone'' like singularities. These results, however, were limited to static geometries, and one would prefer to investigate properties of CFT correlators which correspond to manifestly time-dependent spacetimes, such as gravitational collapse scenarios. Fortunately, it was shown by\cite{hubeny} that this can be achieved by examining the structure of singularities for generic Lorentzian correlators. It is already known that CFT correlators will exhibit light-cone singularities when the operator insertion points are connected by strictly null geodesics. However, by considering boundary-to-boundary null geodesics which penetrate the bulk, the CFT correlator exhibits additional singularities inside the boundary light cone, called\textit{``bulk-cone singularities''}\cite{hubeny}.\\

Through this relationship, one can probe the geometry of certain asymptotically AdS spacetimes given information of boundary data, namely the locus bulk-cone singularities. More specifically, one can attempt to reconstruct the metric of certain classes of spacetimes given such boundary data. This was achieved numerically in\cite{hammer1}, for a general class of static, spherically symmetric, asymptotically AdS $(d+2)$-dimensional spacetimes with metric
\be
\label{genads}
ds^2 = -f(r)\,dt^2+h(r)\,dr^2+r^2\,d\Om^{2}_d.
\ee
In the simplified case where $h(r)=\frac{1}{f(r)}$,\cite{hammer1} was able to extract the metric function $f(r)$ given information of the locus of endpoints of boundary-to-boundary null geodesics encapsulated by the function's worth of information $\D t(\D\p)$. $\D t$ and $\D\p$ are the time and angular separation of null geodesic boundary endpoints respectively. Complementing this work, it was shown in \cite{Bilson} that one can achieve the extraction analytically via the observation that the expression for $\D t(\D\p)$ in terms of $f(r)$ reduces to a standard integral equation with known solution. By applying this solution to various expressions for $\D t(\D\p)$, we were able to recover the same limitation on the extraction as illustrated in \cite{hammer1}, namely that spacetimes admitting null circular orbits with radius $r_m$ can only be recovered in the region $r\in(r_m,\infty)$. Unique to this analytical approach, we showed in \cite{BilsonThesis}, that given $\D t(\D\p)$ as a convergent series, one can always find a convergent series solution for the metric function $f(r)$. This is verified in the case of the AdS-Schwarzschild black hole solution.\\

Pursuing the theme of extracting metric data given information of the dual CFT, one can look for alternative bulk geodesic probes corresponding to some observable in the dual CFT which go behind the region $r\in(r_m,\infty)$. As identified in the review \cite{Hub1}, for fixed energy and angular momentum, boundary spacelike geodesics provide such a probe. Thus, by identifying boundary observables corresponding to such a probe, one may attempt to learn more about the bulk geometry. This was achieved in \cite{Ryu,Ryu1}, which related the area of co-dimension 2 static\footnote{In this paper, we restrict our analysis to static spacetimes where $\pd_t$ is Killing. However one can extend this relationship to time-dependent cases such that the area/entropy formula is fully covariant\cite{Hub}. Extensions to these cases are discussed in \S\ref{discussion}.} minimal surfaces anchored to the boundary, with the entanglement entropy of some subsystem in the boundary CFT. In the specific case of a $2+1$-dimensional bulk, the area of the static minimal surface is the proper length of a zero-energy boundary spacelike geodesic.\\

It had been known for a while\cite{Suss} that black hole entropy $S_{BH}$, which can be calculated from the horizon area $A$ in Planck units using,
\be
S_{BH}=\dfrac{A}{4},
\ee
is related to a quantity known as entanglement entropy (or geometric entropy) of a QFT. This is defined as the von Neumann entropy of a reduced density matrix by tracing out the degrees of freedom of a certain subsystem. Thus it measures how closely entangled a quantum system is. Since the eternal black hole in AdS is related to an entangled state in the CFT\cite{Bal1}, it is natural to assume a relationship between horizon area in the bulk, and entanglement entropy on the boundary. It was this proposal by Ryu \textit{et.al.} that lead to the existence of such a holographic description\footnote{For a good review of all aspects of holographic entanglement entropy, see \cite{Ryu2}.} exist in the form of minimal surface areas. More specifically, it was shown that 
\begin{Con}
\textit{The entanglement entropy $S_A$ of a static subsystem $A$ in a $(d+1)-$dimensional CFT can be determined from a $d$-dimensional static minimal surface $\gamma_A$, in the dual $(d+2)$-dimensional bulk, whose boundary is given by the $(d-1)$-dimensional manifold $\pd\g_A=\pd A$. The entropy is given by applying the usual Berkinstein-Hawking area/entropy relation}
\end{Con}
\be
\label{entent}
S_A=\dfrac{\Area(\g_A)}{4G_N^{(d+2)}}.
\ee
It is this proposal which was used in \cite{hammer2} to numerically recover the metric function $h(r)$ in a $(2+1)$-dimensional bulk described by the metric \eqref{genads}. The boundary observable in this case is the entanglement entropy $S_A(l)$ in an infinitely long one dimensional system where $A$ is an interval of length $l$. This was combined with bulk-cone singularities in null boundary probes to show the one can extract both metric functions in \eqref{genads} in $(2+1)$-dimensions. Following the same reasoning, it was shown in \cite{Bilson} that expressions for $\D t(\D\p)$ and $S_A(l)$ reduce to simple integral equations in terms of the metric functions $f(r)$ and $h(r)$. Thus providing an analytical method for determining metric \eqref{genads}. For spacetimes with an event horizon, \cite{Bilson} showed that $h(r)$ can be extracted in the region $r\in(r_+,\infty)$, where $r_+$ is the horizon radius. Since for all physical spacetimes $r_+\leq r_m$, this method allows one to reconstruct deeper regions of the bulk, thus agreeing with the analysis of \cite{Hub1}.\\

A major limitation of the extraction method in \cite{Bilson}, was the restriction to a $(2+1)-$dimensional bulk spacetime. As is pointed out in \cite{Hub1}, for sensible spacetimes the surfaces anchored to the boundary of higher dimensionality probe deeper into the bulk. So, by not restricting ourselves to 1-dimensional probes, and utilising the full form of \eqref{entent}, one would hope to recover a larger region of the bulk. This is the main motivation behind this paper, where we consider the static, planar symmetric asymptotically AdS metric
\be
ds^2=R^2\left(\dfrac{-h(z)^2\,dt^2+f(z)^2\,dz^2+\sum_{i=1}^d dx_i^2}{z^2}\right),
\ee
and ask whether one can recover the real metric function\footnote{Since we are considering only static surfaces, expressions for the area do not depend on $h(r)$, and so cannot be determined using the method outlined.} $f(z)$ given the entanglement entropy of certain multi-dimensional subsystems of the boundary CFT using \eqref{entent}. This is discussed in the context of two particular simple subsystems defined in \cite{Ryu2}, namely the infinite strip $A_S$ and the circular disk $A_D$. We will show that in the case of the infinite strip, one can again reduce the problem to a known integral equation, thus solving for $f(z)$ in terms of the entanglement entropy $S_{A_S}(l)$. We find the solution consistent with the known pure AdS result and find a series solution for $f(z)$ in perturbed AdS spacetimes. We then turn our attention to the circular disk $A_D$. We find that the reduction in symmetry of the minimal surface equations means that we must resort to a perturbative analysis. We outline a method for determining the area of minimal surfaces anchored to $\pd A_D$ by perturbing around the pure AdS solution of a hemisphere and recover analytical expressions at first order. By doing so, we illustrate in the simple case of the planar black hole, the positive correlation between surface dimension and the depth probed in the bulk.

\section{Static Minimal Surfaces in AdS}

\numberwithin{equation}{section}

In this section, we extend the work of \cite{Bilson} by attempting to extract metric functions in any dimension using the holographic entanglement entropy proposal \eqref{entent}, for certain static subsystems $A$ of the boundary theory. In particular, we consider two specific forms of $A$, the straight belt of width $l$
\be
\label{infstr}
A_S=\{x_i|x_1=x\in[-l/2,l/2],x_{2,3,}\dots_{,m}\in\R\}, 
\ee
and the circular disk of radius $l$
\be
\label{cdisk}
\A_D=\{x_i|r\leq l\},\quad\text{where}\quad r=\sqrt{\sum_{i=1}^{m}x_i^2}.
\ee
Both subsystems are defined in terms of boundary planar spatial coordinates $\{x_i\}\in\R^m$, thus we should look for a class of metrics which respects this planar symmetry on the boundary. A simple class of asymptotically AdS metrics which preserves planar symmetry in the bulk and the boundary is given by
\be
\label{metric}
ds^2=R^2\left(\dfrac{-h(z)^2\,dt^2+f(z)^2\,dz^2+\sum_{i=1}^d dx_i^2}{z^2}\right).
\ee
The metric is written in AdS Poincar\'e coordinates where $z\geq0$ is the bulk coordinate, and where $z=0$ defines the boundary. $f(z)$ and $h(z)$ are arbitrary real functions which have the asymptotic behaviour
\be
h(z)^2,\,\dfrac{1}{f(z)^2}\tend 1+\O(z^{d+1}),\,\,\text{as}\,\,z\tend 0. 
\ee
The most well known example of such a spacetime given by \eqref{metric} is the AdS planar black hole with horizon at $z=z_+$. The metric functions are given by
\be
\label{pBH1}
h(z)^2=\dfrac{1}{f(z)^2}=1-\left(\dfrac{z}{z_+}\right)^{d+1}.
\ee
It can be recovered from the AdSBH in global coordinates by taking the large horizon radius limit $r_+/R\gg1$ and performing the coordinate transformation $z=R^2/r$.

\begin{figure}
\centering
\begin{tabular}{cc}
\begin{minipage}{175pt}
\begin{overpic}[trim = 30mm 0mm 0mm 0mm, clip, width=250pt,height=250pt]{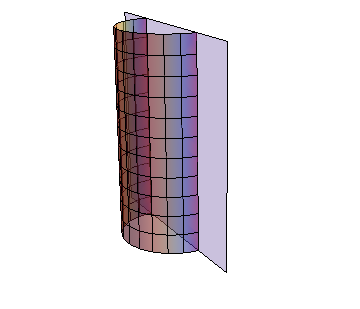}
\thicklines
\put(57,30){\color{black}\vector(0,1){56}}
\put(57,30){\color{black}\vector(0,-1){19}} 
\put(60,50){$L$}
\put(50,92){\color{black}\vector(-4,1){35}}
\put(50,92){\color{black}\vector(4,-1){3}}
\put(30,100){$L$}
\put(30,65){\color{black}\vector(-3,1){7}}
\put(30,65){\color{black}\vector(3,-1){11}}
\put(32,65){$l$}
\put(43,50){$A_S$}
\end{overpic}
\end{minipage}
&
\begin{minipage}{175pt}
\begin{overpic}[trim = 20mm 0mm 0mm 0mm, clip, width=225pt]{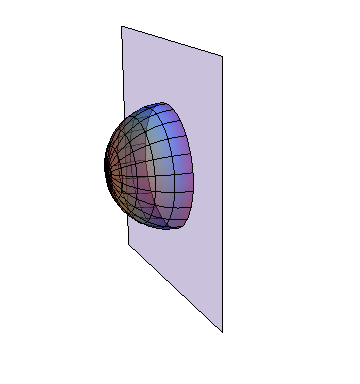}
\thicklines
\put(50,30){\color{black}\vector(0,1){55}}
\put(50,30){\color{black}\vector(0,-1){17}} 
\put(53,50){$L$}
\put(40,90){\color{black}\vector(-3,1){23}}
\put(40,90){\color{black}\vector(3,-1){4}}
\put(30,95){$L$}
\put(29,55){\color{black}\vector(2,-1){7}}
\put(32,50){$l$}
\put(36,60){$A_D$}
\end{overpic}
\end{minipage}
\end{tabular}
\caption{\label{surfplots}These plots depict the shape of static minimal surfaces $\g_A$ anchored to the straight belt $A_S$ of width $l$ and the circular disk $A_D$ of radius $l$. The subsystems $A_S$ and $A_D$ sit on the boundary, where we have regulated the boundary directions to be of length $L$.}
\end{figure}
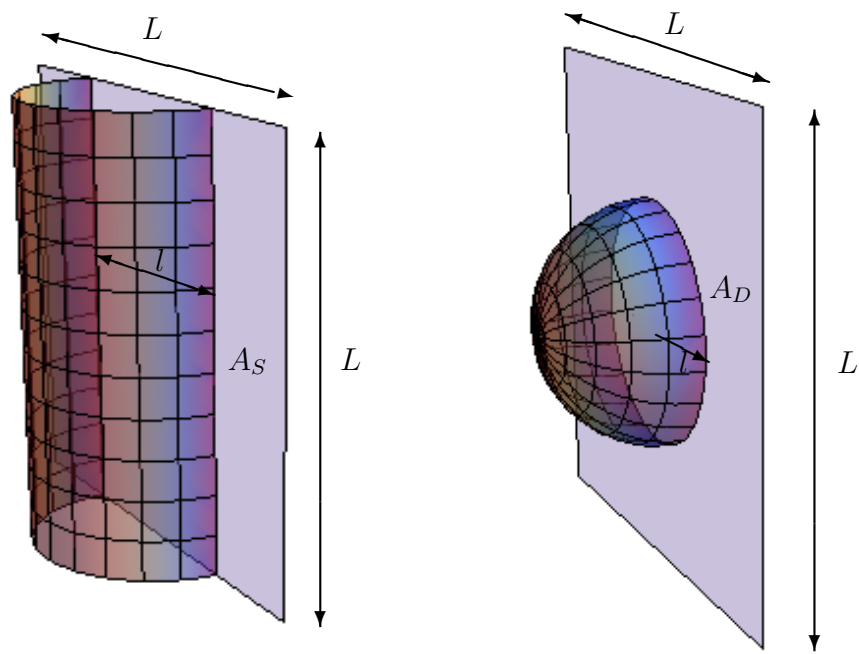

The holographic interpretation of entanglement entropy relates the entanglement entropy $S_A$ in a $d+1$ dimensional boundary CFT to the area of an $d-$dimensional static minimal surface $\g_A$ (see figure \ref{surfplots}) whose boundary is the boundary of the subsystem $A$, i.e. $\pd\g_A=\pd A$. Since the purpose of this chapter is to extract the metric function\footnote{Since we will calculate the area of \textit{static} minimal surfaces, area expressions will not include $h(z)$. Thus one cannot recover $h(z)$ using the method outlined. Although one maybe able to extend such a method by requiring the entanglement entropy be time-dependent \cite{Hub}} $f(z)$ given $S_A$, one must first find an expression for the minimal surface $\g_A$ and thus $\Area(\g_A)$ in terms of $f(z)$. In this section we consider the case of the straight belt $A_S$ defined by \eqref{infstr}.\\

To calculate $\Area(\g_{A_S})$, we need to find the area of a generic surface $N$ such that $\pd N=\pd A_S$. In general, the area of an $m$ dimensional submanifold $N\subset M$ is given by
\be
\label{subarea}
\Area(N)=\int_Nd^mx\sqrt{|g_N|},
\ee
where $g_N=\det(g_{ab})$ and $g_{ab}$ is the pull-back of the metric $\tilde{g}_{\m\n}$ on manifold $M$ to $N$ such that $g_{ab}=\pd_ax^\m\pd_bx^\n\tilde{g}_{\m\n}$.\\

Since $\tilde{g}_{\m\n}$ is given by \eqref{metric}, one can define an embedding function $z=z(x)$ which respects the $\R^{m-1}$ planar symmetry so that $g_{ab}$ is given by
\be
ds_N^2=R^2\left(\dfrac{[z'f(z)]^2dx^2+dx^2+\sum_{i=2}^mdx_i^2}{z^2}\right),\quad\text{where}\,\,z'=\frac{dz}{dx}. 
\ee
Then using \eqref{subarea},
\be
\label{area}
\Area(N_{A_S})=\A_N(l)=R^mL^{m-1}\int_{-\frac{l}{2}}^{\frac{l}{2}}\!dx\,\dfrac{\sqrt{[z'f(z)]^2+1}}{z^m}=R^mL^{m-1}\int_{-\frac{l}{2}}^{\frac{l}{2}}\!\L(z',z,x)\,dx,
\ee
where we have regularized the directions $\{x_2,\cdots,x_m\}$ to be length $L$ (following the same procedure as \cite{Ryu}).\\

To find the minimal surface $\g_{A_S}$, we solve $\delta\A_N(l)=0$. Noting the the Hamiltonian
\be
\H(z,z')=z'\dfrac{d\L}{dz'}-\L=\dfrac{[z'f(z)]^2}{z^m\sqrt{[z'f(z)]^2+1}}-\dfrac{\sqrt{[z'f(z)]^2+1}}{z^m},
\ee
is independent of the variable $x$, we can write $\H$ as a constant such that
\bea
\label{ltz}
\H&=&-z_\ast^{-m},\quad\text{where}\,\,z'|_{z=z_\ast}=0\nn
\thus\dfrac{dz}{dx}&=&\dfrac{\sqrt{z_\ast^{2m}-z^{2m}}}{z^m f(z)}.
\eea
Along with the boundary conditions $z(\pm l/2)=0$, \eqref{ltz} defines the profile function $z(x)$ for the minimal surface $\g_{A_S}$.\\

\begin{figure}[h!t]
\centering
\begin{tabular}{cc}
\begin{minipage}{200pt}
\centerline{\includegraphics[width=200pt]{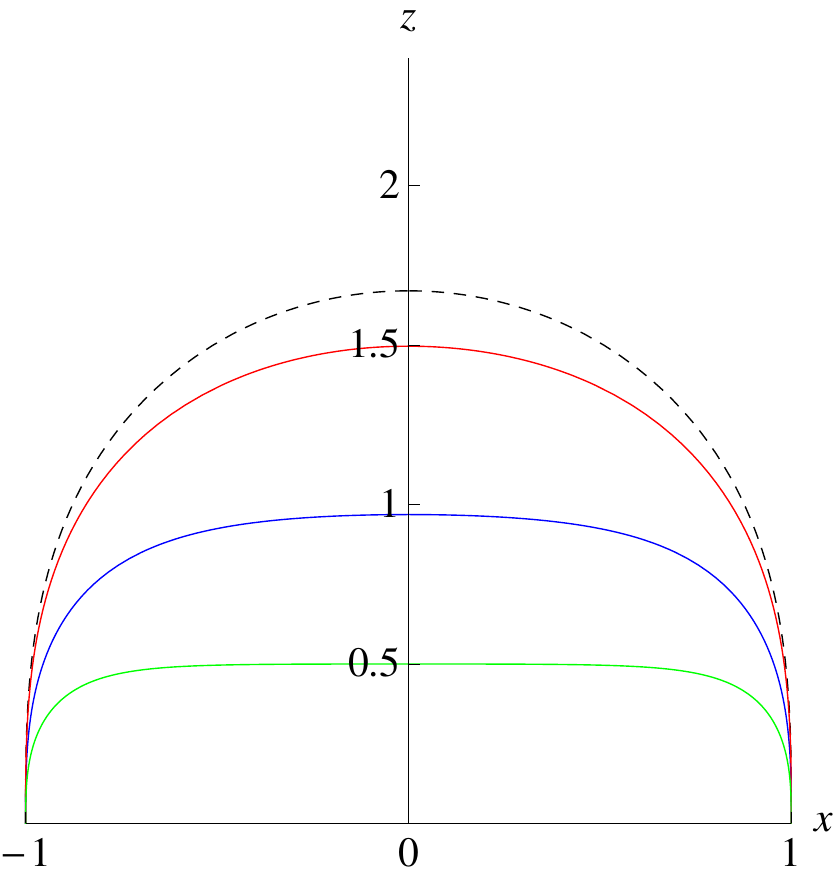}}
\end{minipage}
&
\begin{minipage}{200pt}
\centerline{\includegraphics[width=200pt]{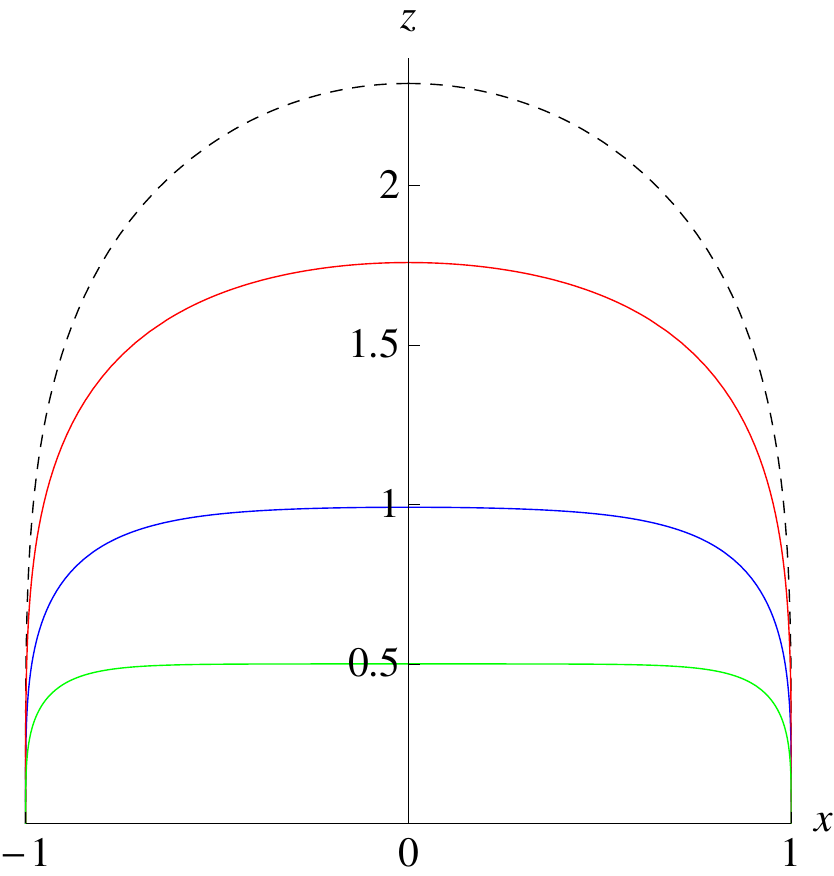}}
\end{minipage}
\end{tabular}
\caption{\label{minsurf}In this figure we plot the profile of static minimal surfaces bounded to a straight belt $A_S$ of width $l=2$ for planar black holes of dimension $4+1$. The left and right figures show minimal surfaces of spacial dimension $2$ and $3$ respectively for different values of horizon depth $z_+=\{0.5(\text{green}),1(\text{blue}),2(\text{red})\}$. The dashed lines correspond to the static minimal surfaces in pure AdS space (i.e. $z_+\tend\infty$).}
\end{figure}

\begin{figure}[!ht]
\centering
\includegraphics[height=3in,width=3in]{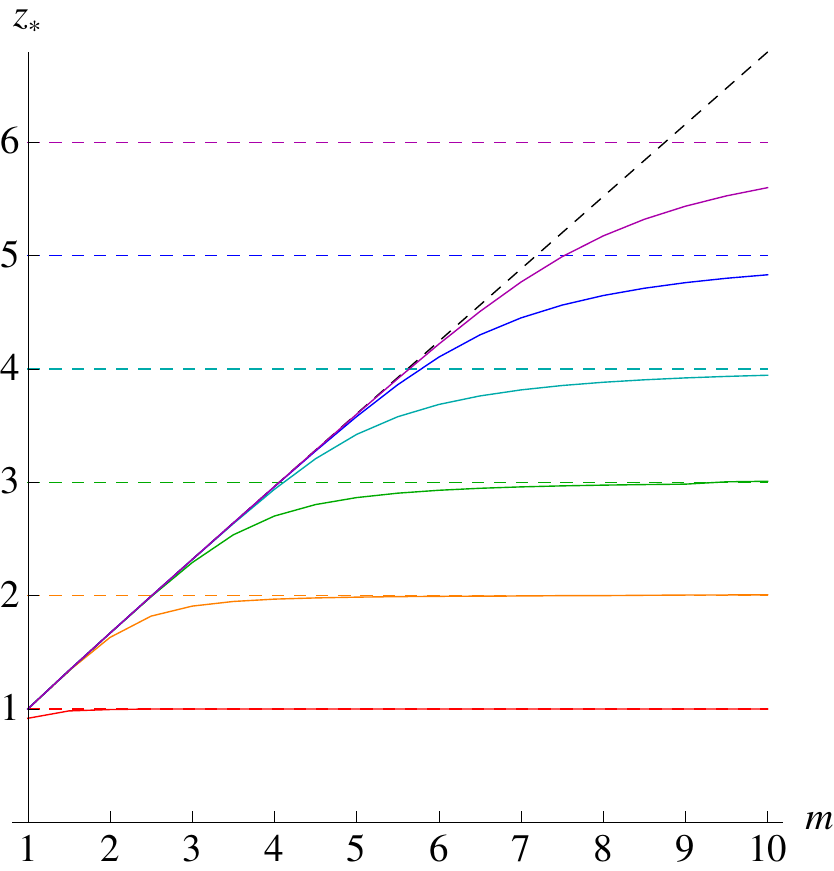}
\caption{\label{mvzplot}This is a plot of $z_\ast$ against the dimension of the minimal surface $m$ in a planar black hole spacetime of dimension $11+1$ for different depths of the horizon plane $z_+$ (shown here by the coloured dashed lines). The pure AdS plot (i.e. $z_+\tend\infty$) given in \eqref{mvz} is shown here by the black dashed line.}
\end{figure}

We illustrate the shapes of $z(x)$ for a $(4+1)-$dimensional planar black hole (see \eqref{pBH1}), with different horizon depths $z_+$, in figure \ref{minsurf}. It is noted that minimal surfaces in pure AdS probe furthest into the bulk. This is because $\g_{A_S}$ has no timelike component, so the presence of the null surface $z=z_+$ flattens the shape of $\g_{A_S}$ such that $z_\ast<z_+$. It is also observed that as one increases the dimension of $A_S$, $z_\ast$ increases. One can show this\footnote{Using $\int_0^1\!dx\,x^{\m-1}(1-x^\la)^{\n-1}=\frac{1}{\la}\frac{\G(\frac{\m}{\la})\G(\n)}{\G(\frac{\m}{\la}+\n)}$} for pure AdS by integrating \eqref{ltz} over the boundary values
\be
\label{lz}
l=2\ds\int_0^{z_\ast}\!dz\,\dfrac{z^m\,f(z)}{\sqrt{z_\ast^{2m}-z^{2m}}},
\ee
and then setting $f(z)=1$,
\bea
\label{mvz}
\za(m)&=&\dfrac{l}{2}\dfrac{\G(\frac{1}{2m})}{\sqrt{\pi}\,\G(\frac{1+m}{2m})}\nn
&\approx&\dfrac{l}{\pi}\left(m-2\g_E-2\psi^{(0)}(1/2)\right).
\eea
where $\g_E$ is the Euler–Mascheroni constant and $\psi^{(0)}$ is the Polygamma function of order zero, and so for pure AdS $z_\ast(m+1)>z_\ast(m)$. The same result is also observed for the planar black hole (see figure \ref{mvzplot}), where the lines \eqref{mvz} and $z_\ast=z_+$ asymptote the curves $z_\ast(m)$ for small and large $m$ respectively. Thus increasing the number of dimensions probes more of the bulk geometry\footnote{Intuitively, this makes sense, since probing extra dimensions costs energy, the minimal surface compensates by moving more into the bulk where distances are smaller.}, and so compared with the previous chapter where we look at a $2+1$ dimensional bulk, we will be able to recover more of the bulk geometry.
 
\newpage

\section{The Method}

We can immediately recover an expression for $\Area(\g_{A_S})$ by substituting \eqref{ltz} into \eqref{area}
\be
\label{marea}
\Area(\g_{A_S})=\A_\g(z_\ast)=2R^mL^{m-1}\int_a^{z_\ast}\!dz\,\dfrac{z_{\ast}^m\,f(z)}{z^m\sqrt{z_\ast^{2m}-z^{2m}}},
\ee
where we have introduced a cut-off surface at $z=a$ close to the boundary to regulate the area functional.\\

We notice that $\A_\g$ is a function of the bulk quantity $z_\ast$, but must be expressed in term of boundary variables only to be considered as boundary data. The entanglement entropy $S_A$ is given as a function of the belt width $l$, and so one must find a map $S_A(l)\mapsto\A_\g(z_\ast)$ which is independent of the metric function $f(z)$. This can be achieved via the chain rule
\be
\label{cr}
\dfrac{dS_A(l)}{dl}=\dfrac{1}{4G_N^{(d+2)}}\dfrac{dz_\ast}{dl}\dfrac{d\A_\g(z_\ast)}{dz_\ast}\Big|_{m=d}.
\ee
Using \eqref{marea} and the Leibniz rule we have,
\be
\label{da}
\dfrac{d\A_\g}{dz_\ast}=2R^mL^{m-1}\left[f(z_\ast)\ds\lim_{z\rightarrow z_\ast}\left\{\dfrac{1}{\sqrt{z_\ast^{2m}-z^{2m}}}\right\}-m\ds\int_0^{z_\ast}\!dz\,\dfrac{z_\ast^{m-1}z^m\,f(z)}{(z_\ast^{2m}-z^{2m})^{\frac{3}{2}}}\right],
\ee
and from \eqref{lz} we have,
\be
\label{zl}
\dfrac{dl}{dz_\ast}=z_\ast^m\left[f(z_\ast)\ds\lim_{z\rightarrow z_\ast}\left\{\dfrac{1}{\sqrt{z_\ast^{2m}-z^{2m}}}\right\}-m\ds\int_0^{z_\ast}\!dz\,\dfrac{z_\ast^{m-1}z^m\,f(z)}{(z_\ast^{2m}-z^{2m})^{\frac{3}{2}}}\right].
\ee
Combining these equations we find that
\be
\label{stoa}
\dfrac{dS_A}{dl}=\dfrac{R^dL^{d-1}}{2z^d_\ast G_N^{(d+2)}}.
\ee
Thus, one can map $l\mapsto z_\ast$ and so $S_A(l)\mapsto\A_\g(z_\ast)$ independently of the metric and we are justified in using \eqref{marea} for the extraction of $f(z)$.\\

By rewriting \eqref{marea} in the more digestible form, where $\tilde{\A}_\g(z_\ast)=\dfrac{\A_\g(z_\ast)}{2R^dL^{d-1}z_\ast^d}$, $p(z)=\dfrac{f(z)}{z^d}$ and $g(z)=z^{2d}$, we have
\be
\tilde{\A}_\g(z_\ast)=\int_a^{z_\ast}\!dz\,\dfrac{p(z)}{\sqrt{g(z_\ast)-g(z)}}.
\ee
The solution to the integral equation of the form $f(x)=\int_a^x\!\frac{y(t)dt}{\sqrt{g(x)-g(t)}}$, where $\frac{dg}{dx}=g'(x)>0$, is given by (\cite{Abel}),
\be
y(x)=\dfrac{1}{\pi}\dfrac{d}{dx}\int_a^x\!dt\,\dfrac{f(t)g'(t)}{\sqrt{g(x)-g(t)}}.
\ee
One can apply this solution to our particular case, and substituting back in for our original variables, we find,
\be
\label{invstr}
f(z)=\dfrac{d\,z^d}{\pi R^dL^{d-1}}\dfrac{d}{dz}\int_a^z\!dz_\ast\,\dfrac{\A_\g(z_\ast)\,z_\ast^{d-1}}{\sqrt{z^{2d}-z_\ast^{2d}}}.
\ee
Thus, using \eqref{stoa} and \eqref{invstr}, we have found a map $S_A(l)\mapsto f(z)$ which is independent of $f(z)$ and so determines the metric function uniquely in the limit $a/z\tend0$.

\section{Checking the Inversion}

We can check the validity of \eqref{invstr} by looking at a simple case where both the entanglement entropy and metric are known, i.e. pure AdS.\\

Using the result from \cite{Ryu}, the entanglement entropy of an $d$ dimensional straight belt of width $l$ defined by \eqref{infstr} is given by\footnote{The $d+2$ dimensional bulk Newton constant $G_N^{(d+2)}$ can be related to boundary CFT parameters via $AdS_{d+2}/CFT_{d+1}$ for a particular choice of boundary theory. For example, when $d=2$, $AdS_4\times S^7$ space in eleven dimensional supergravity is considered to be dual to $2+1$ dimensional $\N=8$ $SU(N)$ SCFT. In this case $\frac{G_N^{(4)}}{R^2}=\frac{3}{2\sqrt{2}}N^{-3/2}$.} 
\be
\label{Ssb}
S_A(l)=\dfrac{2R^d}{4G_N^{(d+2)}(d-1)}\left(\dfrac{L}{a}\right)^{d-1}-\dfrac{2^{d+1}\pi^{d/2}R^d}{4G_N^{(d+2)}(d-1)}\left(\dfrac{\G(\frac{d+1}{2d})}{\G(\frac{1}{2d})}\right)^d\left(\dfrac{L}{l}\right)^{d-1}.
\ee
Using \eqref{stoa}, we find that
\be
\label{lsb}
\dfrac{l}{2}=\za\sqrt{\pi}\dfrac{\G(\frac{d+1}{2d})}{\G(\frac{1}{2d})}.
\ee
Note that this is exactly the result given in \eqref{mvz}, only in this case we have not used the fact that $f(z)=1$.\\
By combining \eqref{Ssb} and \eqref{lsb} with \eqref{entent}, we see that
\be
\label{asb}
\A_S(z_\ast)=\dfrac{2R^d}{d-1}\left(\dfrac{L}{a}\right)^{d-1}-\dfrac{2\sqrt{\pi}R^d}{d-1}\dfrac{\G(\frac{d+1}{2d})}{\G(\frac{1}{2d})}\left(\dfrac{L}{z_\ast}\right)^{d-1}.
\ee
Substituting \eqref{asb} into \eqref{invstr}, we have 
\be
f(z)=\dfrac{2d}{\pi(d-1)}\left(\dfrac{z}{a}\right)^{d-1}I_1(z)-\dfrac{2d\sqrt{\pi}}{\pi(d-1)}\dfrac{\G(\frac{d+1}{2d})}{\G(\frac{1}{2d})}I_2(z),
\ee
where
\bea
I_1(z)&=&z\dfrac{d}{dz}\int^z_adz\dfrac{z^{d-1}_{\ast}}{\sqrt{z^{2d}-z^{2d}_\ast}}\nn
&=&z\dfrac{d}{dz}\int^1_{a/z}dx\dfrac{x^{d-1}}{\sqrt{1-x^{2d}}}\nn
&=&\left(\dfrac{a}{z}\right)^d\left(1+\dfrac{1}{2}\left(\dfrac{a}{z}\right)^{2d}+\O\left(\dfrac{a}{z}\right)^{4d}\right),
\eea
and
\bea
I_2(z)&=&z^d\dfrac{d}{dz}\int^z_a\dfrac{dz_\ast}{\sqrt{z^{2d}-z^{2d}_\ast}}\nn
&=&z^d\dfrac{d}{dz}\dfrac{1}{z^{d-1}}\int^1_{a/z}\dfrac{dx}{\sqrt{1-x^{2d}}}\nn
&=&z^d\dfrac{d}{dz}\dfrac{1}{z^{d-1}}\int^1_0\dfrac{dx}{\sqrt{1-x^{2d}}}-z^m\dfrac{d}{dz}\dfrac{1}{z^{d-1}}\int^{a/z}_0\dfrac{dx}{\sqrt{1-x^{2d}}}\nn
&=&-(d-1)\sqrt{\pi}\dfrac{\G(\frac{2d+1}{2d})}{\G(\frac{d+1}{2d})}+\left(\dfrac{a}{z}\right)\dfrac{1}{\sqrt{1-(a/z)^{2d}}}\nn
&&+(d-1)\int^{a/z}_0\dfrac{dx}{\sqrt{1-x^{2d}}}\nn
&=&-(d-1)\sqrt{\pi}\dfrac{\G(\frac{2d+1}{2d})}{\G(\frac{d+1}{2d})}+d\left(\dfrac{a}{z}\right)\left(1+\O\left(\dfrac{a}{z}\right)^{2d}\right).
\eea
Combining the results, we have
\be
f(z)=1+\dfrac{2d}{\pi(d-1)}\left(\dfrac{a}{z}\right)\left(1-d\sqrt{\pi}\dfrac{\G(\frac{d+1}{2d})}{\G(\frac{1}{2d})}+\O\left(\dfrac{a}{z}\right)^{2d}\right).
\ee
Taking the limit $a/z\rightarrow 0$, we get the required result for pure AdS.

\section{The Series Solution}

One can consider examples of area functions for which one can recover analytic expressions for the metric function $f(z)$ using \eqref{invstr}. Take the infinite series describing a $d-$dimensional surface about the known AdS result \eqref{asb},
\be
\label{pertA}
\A_S(z_\ast)=2R^d\left(\dfrac{L}{z_\ast}\right)^{d-1}\sum_{n=0}^\infty b_n(d)z^n_\ast,
\ee
where $b_0(d)=-\frac{\sqrt{\pi}}{d-1}\frac{\G(\frac{d+1}{2d})}{\G(\frac{1}{2d})}$ and $b_{d-1}(d)=\frac{1}{d-1}\frac{1}{a^{d-1}}$. 

By applying \eqref{invstr}, we have
\bea
f(z)&=&\dfrac{2dz^d}{\pi}\dfrac{d}{dz}\int^z_adz_\ast\dfrac{\sum_{n=0}^\infty b_n(d)z^n_\ast}{\sqrt{z^{2d}-z^{2d}_\ast}}\nn
&=&\dfrac{2dz^d}{\pi}\sum_{n=0}^\infty b_n(d)\dfrac{d}{dz}z^{n+1-d}\int^1_0dx\dfrac{x^n}{\sqrt{1-x^{2d}}}+\O(a/z)\nn
&=&1+\dfrac{1}{\sqrt{\pi}}\sum^\infty_{n=1}b_n(d)(n+1-d)\dfrac{\G(\frac{n+1}{2d})}{\G(\frac{n+d+1}{2d})}z^n\nn
&=&1+\sum^\infty_{n=1}\tilde{b}_nz^n,
\eea
where $\tilde{b}_n=(n+1-d)\frac{\G(\frac{n+1}{2d})}{\G(\frac{n+d+1}{2d})}b_n(d)$.

Using this result, one can recover the planar black hole in $d+2$ dimensions with horizon radius $z_+$. The metric function in this case is given by
\bea
f(z)&=&\dfrac{1}{\sqrt{1-(z/z_+)^{d+1}}}\nn
&=&1+\sum^\infty_{n=1}\dfrac{\G(n+1/2)}{2n!\sqrt{\pi}}\left(\dfrac{z}{z_+}\right)^{n(d+1)}.
\eea
This series is necessarily convergent as $0<z\leq\za<z_+$ (see figure \ref{mvzplot}), and can be recovered using ansatz \eqref{pertA} with
\be
b_{n>0}(d)=
\left\{
	\begin{array}{ll}
		\dfrac{1}{2\sqrt{\pi}(n+1-d)}\dfrac{\G(\frac{2n+1}{2})}{\G(n+1)}\dfrac{\G(\frac{n+d+1}{2d})}{\G(\frac{n+1}{2d})}\dfrac{1}{z_+^n} & \mbox{if } \frac{n}{d+1}\in\Z^+\\
		0 & \mbox{otherwise }
	\end{array}
\right.
\ee
\newpage
\section{Extracting the Bulk Metric for a Circular Disk}

Now we have a method for determining the metric function $f(z)$ in \eqref{metric} in the case of a straight belt $A_S$, we can look at other types of shapes where the calculation of the entanglement entropy in the CFT is relatively simple. In particular we consider the circular disk $A_D$ defined by \eqref{cdisk}.\\

We can rewrite \eqref{metric} in polar coordinates, to respect the symmetry of $A_D$, as
\be
ds^2=R^2\left(\dfrac{-h(z)^2\,dt^2+f(z)^2\,dz^2+dr^2+r^2d\Om^2_{d-1}}{z^2}\right),
\ee
where $r$ is the radial coordinate on the boundary.\\

Following the same procedure as the straight belt, but using the embedding $z=z(r)$, the area of a general $m$ dimensional static surface $N$, such that $\pd N=\pd A_D$, is given by
\bea
\label{aread}
\Area(N)=\A_N(l)&=&R^m\,\text{Vol}(S^{m-1})\int^l_0dr\,r^{m-1}\dfrac{\sqrt{1+(z'f(z))^2}}{z^m}\nn
&=&R^m\,\text{Vol}(S^{m-1})\int^l_0\!\L(z',z,r)\,dr
\eea
Compared to the straight belt, we have $r$ dependence in the Lagrangian and so the Hamiltonian is not constant. Thus using the Euler-Lagrange equation
\be
\dfrac{d}{dr}\dfrac{d\L}{dz'}=\dfrac{d\L}{dz}
\ee
we have
\bea
\label{eomd}
rf(z)^2zz''&+&(m-1)f(z)^4z(z')^3+mrf(z)^2(z')^2\nn
&+&rf(z)f'(z)z(z')^2+(m-1)f(z)^2zz'+mr=0,
\eea
Solving for $z(r)$ with the boundary conditions $z(l)=0$ and $z'(0)=0$ give us the minimal surface $\g_{A_D}$. For pure AdS where $f(z)=1$, the solution to \eqref{eomd} is given by, $z^2+r^2=l^2$. Thus the static minimal surface $\g_{A_D}$ anchored to the boundary of $A_D$, in pure AdS space, is a semi-circle of radius $l$ for $m=1$ and a hemisphere of radius $l$ for $m>2$.

\subsection{A Perturbative Approach}

A general solution to \eqref{eomd} is intractable, but we can attempt to find $z(r)$ perturbatively, solving order by order in a parameter $\e$ about the known AdS solution.\\

To proceed, we will find the perturbation equations simplify if we rewrite \eqref{eomd} in the coordinate $v=r^2$, giving
\bea
\label{eomd2}
2f(z)^2[mz'(z+2vz')+2vzz'']&+&8(m-1)vf(z)^4z(z')^3\nn
&+&4vf(z)zf'(z)(z')^2+m=0
\eea
where $z'=\frac{dz}{dv}$ etc.

Now we can ansatz a solution for $z(v)$ in orders of $\e$, which will eventually be set to 1,
\be
\label{antz}
f(z)=1+\sum_{i=1}^\infty\e^ia_iz^{i(d+1)}\quad\text{and}\quad z(v)=\sqrt{b-v}+\sum_{i=1}^\infty\e^iz_{(i)}(v),
\ee
where $v=r^2$ and $b=l^2$.\\

Substituting ansatz \eqref{antz} into \eqref{eomd2}, the $\O(\e^n)$ equation is given by
\be
\label{peom}
A(v)\,z''_{(n)}(v)+B(v)\,z'_{(n)}(v)+C(v)\,z_{(n)}(v)=D_{(n-1)}(v),
\ee
where
\be
A(v)=4v\sqrt{b-v},\,\,B(v)=\dfrac{2(bm-3v)}{\sqrt{b-v}},\,\,C(v)=-\dfrac{bm}{(b-v)^{3/2}},
\ee
and $D_{(n-1)}(v)$ is determined from the $(n-1)^\text{th}$ solution.\\

The homogeneous equation has a non-trivial particular solution, 
\be
z^{(1)}(v)=(b-v)^{-1/2}.
\ee
From this one can construct the second solution \cite{Diff}
\bea
z^{(2)}(v)&=&z^{(1)}(v)\int\dfrac{W(v)}{(z^{(1)}(v))^2}\,dv,\quad\text{where}\quad W(v)=e^{-\int\frac{B(v)}{A(v)}\,dv}\nn
&=&\dfrac{1}{\sqrt{b-v}}\int dv\dfrac{(b-v)^{\frac{5-m}{2}}}{v^{m/2}}.
\eea
The general solution of \eqref{peom} is then given by,
\bea
\label{pmin}
z_{(n)}(v)=C_1z^{(1)}(v)&+&C_2z^{(2)}(v)+z^{(2)}(v)\int\!dv\dfrac{D_{(n-1)}(v)\,z^{(1)}(v)}{A(v)\,W(v)}\nn
&-&z^{(1)}(v)\int\!dv\dfrac{D_{(n-1)}(v)\,z^{(2)}(v)}{A(v)\,W(v)},
\eea
where $C_1,C_2$ are determined by the boundary conditions $z_{(n)}(b)=0$ and $\lim_{v\tend0}(\sqrt{v}z_{(n)}'(v))=0$. We see immediately that as $v\tend0$, $\sqrt{v}z'^{(2)}(v)\sim\O(v^{-\frac{m-1}{2}})$, and so we must set $C_2=0$. Expression \eqref{pmin} then becomes
\bea
\label{pmin2}
z_{(n)}(v)&=&\dfrac{C_1}{\sqrt{b-v}}+\sqrt{b-v}\left(\int dv\dfrac{(b-v)^{\frac{m-1}{2}}}{v^{m/2}}\right)\int\!dv\,D_{(n-1)}(v)\dfrac{v^{\frac{m-2}{2}}}{4(b-v)^{\frac{m+1}{2}}}\nn
&&-\sqrt{b-v}\int\!dv\left(D_{(n-1)}(v)\dfrac{v^{\frac{m-2}{2}}}{4(b-v)^{\frac{m+1}{2}}}\int dv\dfrac{(b-v)^{\frac{m-1}{2}}}{v^{m/2}}\right).
\eea
The constant $C_1$ is determined by the condition that $z_{(n)}(v)$ is non-singular at the boundary $v=b$.\\
 
\begin{figure}[h!t]
\centering
\begin{tabular}{cc}
\begin{minipage}{200pt}
\centerline{\includegraphics[width=200pt]{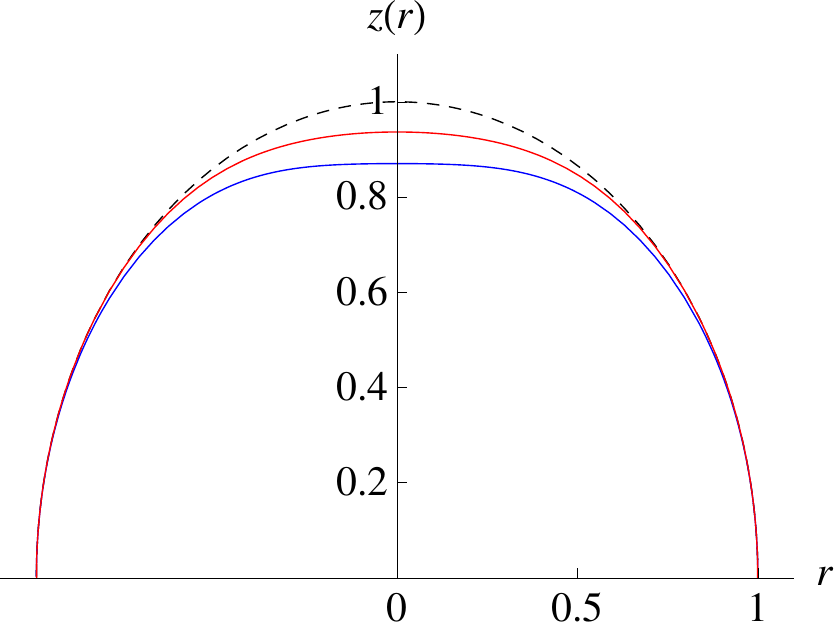}}
\end{minipage}
&
\begin{minipage}{200pt}
\centerline{\includegraphics[width=200pt]{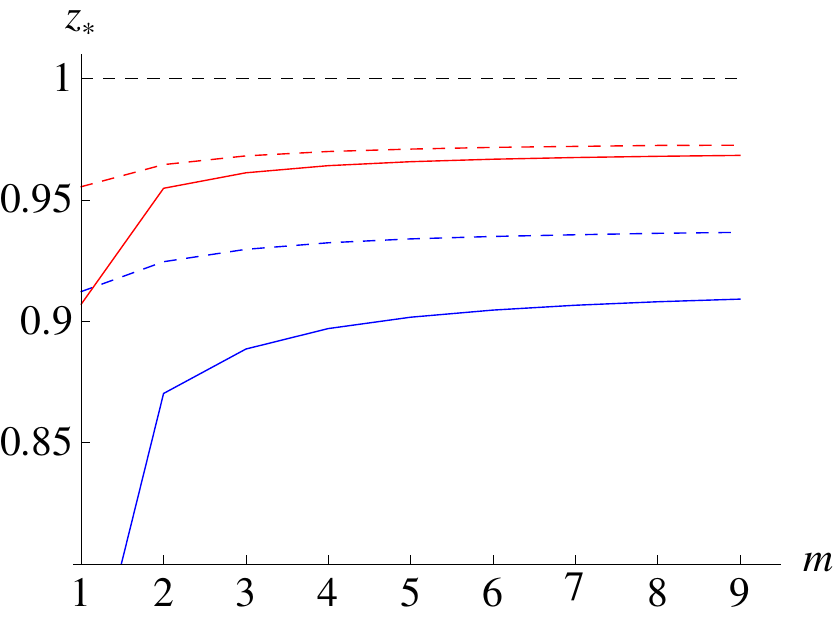}}
\end{minipage}
\end{tabular}
\caption{\label{minsurfdisk}The left diagram plots profile functions $z(r)$ anchored to the the circular disk $A_D$ of radius $l=1$ and dimension $m$, for the metric function $f(z)=(1-z^{10})^{-1/2}$. This spacetime describes a planar black hole in $10+1$ dimensions of horizon depth $z_+=1$. The dashed, red and blue curves show pure AdS, the numerical solution and the first order perturbed solution to \eqref{eomd} respectively for $m=2$. The $r$ axis is the boundary of the spacetime. The right diagram shows how the maximum height $\za$ changes for these curves as one increases the dimension of the circular disk up to the maximum allowed by the dimension of the spacetime.}
\end{figure}

The integrals is \eqref{pmin2} can be solved analytically at first order given the dimension of the surface $m$ and $D_{(0)}=\frac{2bm-3v}{\sqrt{b-v}}a_1$. In figure \ref{minsurfdisk} we compare the shape of a 2-dimensional minimal surface $\g_{A_D}$ using the perturbative method to first order with the numerical solution for a particular planar black hole. We also compare the numerical and first order solutions for increasing $m$. We note the same result as for the straight belt that $\za(m+1)>\za(m)$, where $z_\ast=z_{(1)}(0)$ is the maximum distance the minimal surface probes the bulk. We also find the expected pattern that the perturbative solution approaches the numerical solution as it nears the pure AdS hemisphere.  

In theory, we have an expression for the minimal surface to any order in the perturbation, we can simply plug \eqref{pmin} and \eqref{antz} into \eqref{aread} and equate orders of $\e$ to determine the perturbed minimal surface area $\Area(\g_{A_D})=\A_\g(l)$. Computing $\A_\g(l)$ at first order in $\e$ by substituting \eqref{antz} into \eqref{aread} with $r=l\sqrt{1-y^2}$, we have
\bea
\A_\g(l)&=&R^m\,\text{Vol}(S^{m-1})\int^1_{a/l}dy\dfrac{(1-y^2)^{\frac{m-2}{2}}}{y^m}\nn
&&+R^m\,\text{Vol}(S^{m-1})\dfrac{\e a_1}{l}\dfrac{d}{2}\int^1_{a/l}dy\dfrac{(1-y^2)^{\frac{m-2}{2}}}{y^{m+2}}z_{(1)}(y)\nn
&&-R^m\,\text{Vol}(S^{m-1})\dfrac{\e a_1}{l}\int^1_{a/l}dy\dfrac{(1-y^2)^{m/2}}{y^m}z'_{(1)}(y)
\eea
One can now relate the expressions for $\A_\g(l)$ and $f(z)$ through the parameter $a_1$ once $\e$ is set to 1. 
\newpage
\section{Discussion}
\label{discussion}

In this paper, we extended the ideas and methods of metric extraction explored in \cite{Bilson} to another set of bulk probes anchored to the boundary, namely multi-dimensional minimal surfaces. As discussed in \cite{Hub1,Hubun}, higher-dimensional surfaces probe deeper for spacetimes satisfying energy conditions, so it makes sense to study such probes. We considered static, $(d+2)-$dimensional planar symmetric, asymptotically AdS spacetimes of the form\footnote{We use the tilde notation to distinguish from the metric functions described earlier in this chapter.}
\be
\label{plmet}
ds^2=R^2\left(\dfrac{-\tilde{h}(z)^2\,dt^2+\tilde{f}(z)^2\,dz^2+d\Sigma^2_d}{z^2}\right).  
\ee
We pick this particular form of the metric since this is the natural generalisation of the pure AdS metric ($\tilde{h}(z)=\tilde{f}(z)=1$) used in \cite{Ryu,Ryu1}, where analytical expressions for the entanglement entropy were found. Since the minimal surfaces must be anchored to some shape on the boundary, we picked two cases. The straight belt given defined in \eqref{infstr} where $d\Sigma^2_d=\sum_{i=1}^d dx_i^2$, and the circular disk defined in \eqref{cdisk} where $d\Sigma^2_d=dr^2+r^2d\Om^2_{d-1}$.\\

We first considered the straight belt $A_S$. Comparing minimal surfaces of different dimension at fixed width on the boundary, we found for pure AdS, and planar black hole geometries, that higher dimensional surfaces probe further into the bulk, thus agreeing with \cite{Hub1}. We then found and expression for the area of these minimal surfaces $\A_{\g}(z_\ast)$ in terms of the metric function $\tilde{f}(z)$. We showed that the function $\A_{\g}(z_\ast)$ can be determined from the entanglement entropy $S_{A_S}(l)$ of the straight belt $A_S$ of width $l$ in the boundary dual CFT without prior knowledge of the bulk geometry. Via the observation that equation \eqref{marea} was an integral equation with known solution \eqref{invstr}, we were able to extract the metric function $\tilde{f}(z)$, with knowledge of the entanglement entropy $S_{A_S}(l)$ of the straight belt $A_S$ of width $l$. This was confirmed for the known pure AdS result. We were also able to find a series solution for $\tilde{f}(z)$, where one was able to recover the planar black hole.\\

One may ask the question whether one can extract $\tilde{h}(z)$? The answer is negative if one considers only static surfaces, since the area functional does not depend on the timelike component of the metric. However, if we make the straight belt time-dependent, then using the covariant holographic entanglement entropy proposal \cite{Hub}, one might be able to determine $\tilde{h}(z)$ using extremal surfaces.\\

We then turned our attention to the circular disk shape $A_D$. One finds in this case that the minimal surface equations are not as simple, since the area density (see \eqref{aread}) depends on the boundary coordinate $r$. As such, one is forced to consider a perturbative approach\footnote{A Hamilton-Jacobi method was also attempted. In this case, the Hamilton-Jacobi equation becomes
\be
\left(\dfrac{\pd S}{\pd r}\right)^2+\dfrac{1}{\tilde{f}(z)^2}\left(\dfrac{\pd S}{\pd z}\right)^2=\dfrac{r^{2m-2}}{z^{2m}},
\ee
where $S=S(z,r,C)$ is Hamilton's principle function. If this can be solved for $S$, one can determine $z'(r)$ via the equation $\frac{\pd\L}{\pd z'}=\frac{\pd S}{\pd z}$, where $\L=\L(z,z',r)$ is given in \eqref{area}.}. This was achieved by perturbing about the known pure AdS solution where the minimal surface is a $d-$dimensional hemisphere of radius $l$, given in terms of the profile function $z(r)$ as $z^2+r^2=l^2$. Fortunately, the minimal surface equations reduce to non-linear second order, inhomogeneous differential equations with a known particular solution \eqref{peom}. This allowed us to construct a general solution at each order in the perturbation, and at first order, find an analytical expression for the minimal surface profile function $z(r)$ in any dimension. One could then compare such a solution to the solution obtained by numerically solving the equations of motion. This was illustrated in Figure \ref{minsurfdisk}, where it was observed that as the numerical solution approaches the pure AdS solution, the first order perturbative solution becomes more accurate. Obtaining higher order terms to the profile function proved intractable since the integrals of \eqref{pmin} cannot be solved analytically, and a series solution to the integrals ran into convergence issues.\\

How could one extend the methods presented in this paper to other cases? The most natural extension is to consider other shapes on the boundary where expressions for the entanglement entropy dual to pure AdS are known. A known example\footnote{See \cite{Ryu3} for a holographic calculation of the entanglement entropy in pure AdS} is the cusp $A_W$ in a $(2+1)-$dimensional CFT. In polar boundary coordinates $(r,\th)$, a cusp of angle $\Om$ is defined in terms of its boundary
\be
\pd A_W=\{(r,\th)|0\leq r<\infty,\th=0\}\cup\{(r,\th)|0\leq r<\infty,\th=\Om\}.
\ee
One could then attempt to solve for the area of the minimal surface anchored to $\pd A_W$ in terms of the metric function $\tilde{f}(z)$. However, one would have to find a suitable regulator for the area of the cusp in the bulk.

\newpage
\bibliographystyle{utphys}
\bibliography{paper2}

\providecommand{\href}[2]{#2}\begingroup\raggedright\begin{thebibliography}{10}

\bibitem{Maldacena}
J.~M. Maldacena, ``{The large N limit of superconformal field theories and
  supergravity},'' \href{http://dx.doi.org/10.1023/A:1026654312961}{{\em Adv.
  Theor. Math. Phys.} {\bfseries 2} (1998) 231--252},
\href{http://arxiv.org/abs/hep-th/9711200}{{\ttfamily arXiv:hep-th/9711200}}.

\bibitem{MAGOO}
O.~Aharony, S.~S. Gubser, J.~M. Maldacena, H.~Ooguri, and Y.~Oz, ``{Large N
  field theories, string theory and gravity},''
  \href{http://dx.doi.org/10.1016/S0370-1573(99)00083-6}{{\em Phys. Rept.}
  {\bfseries 323} (2000) 183--386},
\href{http://arxiv.org/abs/hep-th/9905111}{{\ttfamily arXiv:hep-th/9905111}}.

\bibitem{Witten1}
E.~Witten, ``{Anti-de Sitter space and holography},'' {\em Adv. Theor. Math.
  Phys.} {\bfseries 2} (1998) 253--291,
\href{http://arxiv.org/abs/hep-th/9802150}{{\ttfamily arXiv:hep-th/9802150}}.

\bibitem{Hol5}
L.~Susskind and E.~Witten, ``{The holographic bound in anti-de Sitter space},''
\href{http://arxiv.org/abs/hep-th/9805114}{{\ttfamily arXiv:hep-th/9805114}}.

\bibitem{adscft2}
G.~T. Horowitz and J.~Polchinski, ``{Gauge / gravity duality},''
\href{http://arxiv.org/abs/gr-qc/0602037}{{\ttfamily arXiv:gr-qc/0602037}}.

\bibitem{largeN}
G.~'t~Hooft, ``{A PLANAR DIAGRAM THEORY FOR STRONG INTERACTIONS},''
\href{http://dx.doi.org/10.1016/0550-3213(74)90154-0}{{\em Nucl. Phys.}
  {\bfseries B72} (1974) 461}.

\bibitem{HR}
K.~Skenderis, ``{Lecture notes on holographic renormalization},''
  \href{http://dx.doi.org/10.1088/0264-9381/19/22/306}{{\em Class. Quant.
  Grav.} {\bfseries 19} (2002) 5849--5876},
\href{http://arxiv.org/abs/hep-th/0209067}{{\ttfamily arXiv:hep-th/0209067}}.

\bibitem{Tmn}
V.~Balasubramanian and P.~Kraus, ``{A stress tensor for anti-de Sitter
  gravity},'' \href{http://dx.doi.org/10.1007/s002200050764}{{\em Commun. Math.
  Phys.} {\bfseries 208} (1999) 413--428},
\href{http://arxiv.org/abs/hep-th/9902121}{{\ttfamily arXiv:hep-th/9902121}}.

\bibitem{Witten}
E.~Witten, ``{Anti-de Sitter space, thermal phase transition, and confinement
  in gauge theories},'' {\em Adv. Theor. Math. Phys.} {\bfseries 2} (1998)
  505--532,
\href{http://arxiv.org/abs/hep-th/9803131}{{\ttfamily arXiv:hep-th/9803131}}.

\bibitem{Maldacena1}
J.~M. Maldacena, ``{Eternal black holes in Anti-de-Sitter},'' {\em JHEP}
  {\bfseries 04} (2003) 021,
\href{http://arxiv.org/abs/hep-th/0106112}{{\ttfamily arXiv:hep-th/0106112}}.

\bibitem{AdS/QCD1}
K.~Peeters and M.~Zamaklar, ``{The string/gauge theory correspondence in
  QCD},'' \href{http://dx.doi.org/10.1140/epjst/e2007-00379-0}{{\em Eur. Phys.
  J. ST} {\bfseries 152} (2007) 113--138},
\href{http://arxiv.org/abs/0708.1502}{{\ttfamily arXiv:0708.1502 [hep-ph]}}.

\bibitem{AdS/QCD2}
D.~Mateos, ``{String Theory and Quantum Chromodynamics},''
  \href{http://dx.doi.org/10.1088/0264-9381/24/21/S01}{{\em Class. Quant.
  Grav.} {\bfseries 24} (2007) S713--S740},
\href{http://arxiv.org/abs/0709.1523}{{\ttfamily arXiv:0709.1523 [hep-th]}}.

\bibitem{AdS/QCD3}
S.~S. Gubser and A.~Karch, ``{From gauge-string duality to strong interactions:
  a Pedestrian's Guide},''
  \href{http://dx.doi.org/10.1146/annurev.nucl.010909.083602}{{\em Ann. Rev.
  Nucl. Part. Sci.} {\bfseries 59} (2009) 145--168},
\href{http://arxiv.org/abs/0901.0935}{{\ttfamily arXiv:0901.0935 [hep-th]}}.

\bibitem{AdS/QCD4}
S.~S. Gubser, ``{Heavy ion collisions and black hole dynamics},''
\href{http://dx.doi.org/10.1007/s10714-007-0473-8}{{\em Gen. Rel. Grav.}
  {\bfseries 39} (2007) 1533--1538}.

\bibitem{AdS/CMT1}
J.~McGreevy, ``{Holographic duality with a view toward many-body physics},''
  \href{http://dx.doi.org/10.1155/2010/723105}{{\em Adv. High Energy Phys.}
  {\bfseries 2010} (2010) 723105},
\href{http://arxiv.org/abs/0909.0518}{{\ttfamily arXiv:0909.0518 [hep-th]}}.

\bibitem{AdS/CMT2}
S.~A. Hartnoll, ``{Lectures on holographic methods for condensed matter
  physics},'' \href{http://dx.doi.org/10.1088/0264-9381/26/22/224002}{{\em
  Class. Quant. Grav.} {\bfseries 26} (2009) 224002},
\href{http://arxiv.org/abs/0903.3246}{{\ttfamily arXiv:0903.3246 [hep-th]}}.

\bibitem{AdS/CMT3}
C.~P. Herzog, ``{Lectures on Holographic Superfluidity and
  Superconductivity},''
  \href{http://dx.doi.org/10.1088/1751-8113/42/34/343001}{{\em J. Phys.}
  {\bfseries A42} (2009) 343001},
\href{http://arxiv.org/abs/0904.1975}{{\ttfamily arXiv:0904.1975 [hep-th]}}.

\bibitem{AdS/CMT4}
G.~T. Horowitz, ``{Introduction to Holographic Superconductors},''
\href{http://arxiv.org/abs/1002.1722}{{\ttfamily arXiv:1002.1722 [hep-th]}}.

\bibitem{Fid}
L.~Fidkowski, V.~Hubeny, M.~Kleban, and S.~Shenker, ``{The black hole
  singularity in AdS/CFT},'' {\em JHEP} {\bfseries 02} (2004) 014,
\href{http://arxiv.org/abs/hep-th/0306170}{{\ttfamily arXiv:hep-th/0306170}}.

\bibitem{Bal}
V.~Balasubramanian and S.~F. Ross, ``{Holographic particle detection},''
  \href{http://dx.doi.org/10.1103/PhysRevD.61.044007}{{\em Phys. Rev.}
  {\bfseries D61} (2000) 044007},
\href{http://arxiv.org/abs/hep-th/9906226}{{\ttfamily arXiv:hep-th/9906226}}.

\bibitem{Louko}
J.~Louko, D.~Marolf, and S.~F. Ross, ``{On geodesic propagators and black hole
  holography},'' \href{http://dx.doi.org/10.1103/PhysRevD.62.044041}{{\em Phys.
  Rev.} {\bfseries D62} (2000) 044041},
\href{http://arxiv.org/abs/hep-th/0002111}{{\ttfamily arXiv:hep-th/0002111}}.

\bibitem{Kraus}
P.~Kraus, H.~Ooguri, and S.~Shenker, ``{Inside the horizon with AdS/CFT},''
  \href{http://dx.doi.org/10.1103/PhysRevD.67.124022}{{\em Phys. Rev.}
  {\bfseries D67} (2003) 124022},
\href{http://arxiv.org/abs/hep-th/0212277}{{\ttfamily arXiv:hep-th/0212277}}.

\bibitem{Fest}
G.~Festuccia and H.~Liu, ``{Excursions beyond the horizon: Black hole
  singularities in Yang-Mills theories. I},'' {\em JHEP} {\bfseries 04} (2006)
  044,
\href{http://arxiv.org/abs/hep-th/0506202}{{\ttfamily arXiv:hep-th/0506202}}.

\bibitem{Freiv}
B.~Freivogel {\em et al.}, ``{Inflation in AdS/CFT},'' {\em JHEP} {\bfseries
  03} (2006) 007,
\href{http://arxiv.org/abs/hep-th/0510046}{{\ttfamily arXiv:hep-th/0510046}}.

\bibitem{hubeny}
V.~E. Hubeny, H.~Liu, and M.~Rangamani, ``{Bulk-cone singularities and
  signatures of horizon formation in AdS/CFT},'' {\em JHEP} {\bfseries 01}
  (2007) 009,
\href{http://arxiv.org/abs/hep-th/0610041}{{\ttfamily arXiv:hep-th/0610041}}.

\bibitem{hammer1}
J.~Hammersley, ``{Extracting the bulk metric from boundary information in
  asymptotically AdS spacetimes},'' {\em JHEP} {\bfseries 12} (2006) 047,
\href{http://arxiv.org/abs/hep-th/0609202}{{\ttfamily arXiv:hep-th/0609202}}.

\bibitem{Bilson}
S.~Bilson, ``{Extracting spacetimes using the AdS/CFT conjecture},''
  \href{http://dx.doi.org/10.1088/1126-6708/2008/08/073}{{\em JHEP} {\bfseries
  08} (2008) 073},
\href{http://arxiv.org/abs/0807.3695}{{\ttfamily arXiv:0807.3695 [hep-th]}}.

\bibitem{BilsonThesis}
S.~Bilson, {\em Gauge/Graviy Duality: Recovering the Bulk from the Boundary
  using AdS/CFT}.
\newblock {PhD} in {E}lementaty {P}article {T}heory, Depatment of Mathematical
  Sciences -- Durham University, 2010.

\bibitem{Hub1}
V.~E. Hubeny and M.~Rangamani, ``{A holographic view on physics out of
  equilibrium},''
\href{http://arxiv.org/abs/1006.3675}{{\ttfamily arXiv:1006.3675 [hep-th]}}.

\bibitem{Ryu}
S.~Ryu and T.~Takayanagi, ``{Aspects of holographic entanglement entropy},''
  {\em JHEP} {\bfseries 08} (2006) 045,
\href{http://arxiv.org/abs/hep-th/0605073}{{\ttfamily arXiv:hep-th/0605073}}.

\bibitem{Ryu1}
S.~Ryu and T.~Takayanagi, ``{Holographic derivation of entanglement entropy
  from AdS/CFT},'' \href{http://dx.doi.org/10.1103/PhysRevLett.96.181602}{{\em
  Phys. Rev. Lett.} {\bfseries 96} (2006) 181602},
\href{http://arxiv.org/abs/hep-th/0603001}{{\ttfamily arXiv:hep-th/0603001}}.

\bibitem{Hub}
V.~E. Hubeny, M.~Rangamani, and T.~Takayanagi, ``{A covariant holographic
  entanglement entropy proposal},''
  \href{http://dx.doi.org/10.1088/1126-6708/2007/07/062}{{\em JHEP} {\bfseries
  07} (2007) 062},
\href{http://arxiv.org/abs/0705.0016}{{\ttfamily arXiv:0705.0016 [hep-th]}}.

\bibitem{Suss}
L.~Susskind and J.~Uglum, ``{Black hole entropy in canonical quantum gravity
  and superstring theory},''
  \href{http://dx.doi.org/10.1103/PhysRevD.50.2700}{{\em Phys. Rev.} {\bfseries
  D50} (1994) 2700--2711},
\href{http://arxiv.org/abs/hep-th/9401070}{{\ttfamily arXiv:hep-th/9401070}}.

\bibitem{Bal1}
V.~Balasubramanian, P.~Kraus, A.~E. Lawrence, and S.~P. Trivedi, ``{Holographic
  probes of anti-de Sitter space-times},''
  \href{http://dx.doi.org/10.1103/PhysRevD.59.104021}{{\em Phys. Rev.}
  {\bfseries D59} (1999) 104021},
\href{http://arxiv.org/abs/hep-th/9808017}{{\ttfamily arXiv:hep-th/9808017}}.

\bibitem{Ryu2}
T.~Nishioka, S.~Ryu, and T.~Takayanagi, ``{Holographic Entanglement Entropy: An
  Overview},'' \href{http://dx.doi.org/10.1088/1751-8113/42/50/504008}{{\em J.
  Phys.} {\bfseries A42} (2009) 504008},
\href{http://arxiv.org/abs/0905.0932}{{\ttfamily arXiv:0905.0932 [hep-th]}}.

\bibitem{hammer2}
J.~Hammersley, ``{Numerical metric extraction in AdS/CFT},''
  \href{http://dx.doi.org/10.1007/s10714-007-0564-6}{{\em Gen. Rel. Grav.}
  {\bfseries 40} (2008) 1619--1652},
\href{http://arxiv.org/abs/0705.0159}{{\ttfamily arXiv:0705.0159 [hep-th]}}.

\bibitem{Abel}
A.~Polyanin and A.~Manzhirov, {\em Handbook of Integral Equations}.
\newblock CRC Press LLC, 2000.

\bibitem{Diff}
A.~Polyanin and V.~Zaitsev, {\em Handbook of Exact Solutions for Ordinary
  Differential Equations}.
\newblock CRC Press LLC, 1995.

\bibitem{Hubun}
V.~E. Hubeny, ``{in preperation}.''.

\bibitem{Ryu3}
T.~Hirata and T.~Takayanagi, ``{AdS/CFT and strong subadditivity of
  entanglement entropy},'' {\em JHEP} {\bfseries 02} (2007) 042,
\href{http://arxiv.org/abs/hep-th/0608213}{{\ttfamily arXiv:hep-th/0608213}}.

\end{thebibliography}\endgroup
   
\end{document}